Article type: Research Article

Title:
**NV Center Electron Paramagnetic Resonance of a Single Nanodiamond Attached to an Individual Biomolecule**

*Richelle M. Teeling-Smith, Young Woo Jung, Nicolas Scozzaro, Jeremy Cardellino, Isaac Rampersaud, Justin A. North, Marek Šimon, Vidya P. Bhallamudi, Arfaan Rampersaud, Ezekiel Johnston-Halperin\*, Michael G. Poirier\*, P. Chris Hammel\**

Dr. R. M. Teeling-Smith, N. Scozzaro, J. Cardellino, Dr. J. A. North, Dr. M. Šimon, Dr. V. P. Bhallamudi, Prof. E. Johnston-Halperin\*, Prof. M. G. Poirier\*, Prof. P. C. Hammel\*
Department of Physics,
The Ohio State University,
Columbus, OH 43210, USA
E-mail: ejh@physics.osu.edu, poirier.18@osu.edu, hammel@physics.osu.edu

Dr. Y. W. Jung
Samsung Electronics,
San #24 Nongseo-Dong, Giheung-Gu,
Yongin-City, Gyronggi-Do 446-711, Korea

Dr. A. Rampersaud, I. Rampersaud
Columbus Nanoworks,
1507 Chambers Rd., Columbus, OH, 43210, USA

A key limitation of electron paramagnetic resonance (EPR), an established and powerful tool for studying atomic-scale biomolecular structure and dynamics is its poor sensitivity; samples containing in excess of $10^{12}$ labeled biomolecules are required in typical experiments. In contrast, single molecule measurements provide improved insights into heterogeneous behaviors that can be masked by ensemble measurements and are often essential for illuminating the molecular mechanisms behind the function of a biomolecule. We report EPR measurements of a single labeled biomolecule that merge these two powerful techniques. We selectively label an individual double-stranded DNA molecule with a single nanodiamond containing nitrogen-vacancy (NV) centers, and optically detect the paramagnetic resonance of NV spins in the nanodiamond probe. Analysis of the spectrum reveals that the nanodiamond probe has complete rotational freedom and that the characteristic time scale for reorientation of the nanodiamond probe is slow compared to the transverse spin relaxation time. This demonstration of EPR spectroscopy of a single nanodiamond labeled DNA provides the foundation for the development of single molecule magnetic resonance studies of complex biomolecular systems.

## 1. Introduction

Single molecule studies are essential to understanding the structural dynamics of biomolecules and the mechanisms responsible for their functions.[1-10] Many single molecule techniques probe local biomolecular dynamics[1-5, 7, 8, 10] and their impact on distributions of structures and correlation times [1, 3-5, 8, 10] of these systems, providing insight into mechanisms underlying biomolecular functionality [1, 2, 4, 6, 8-10]. Single molecule techniques typically rely on the attachment of a probe that can be detected and/or manipulated at the single probe level, such as organic fluorophores, quantum dots, gold nanoparticles, super paramagnetic particles and polystyrene beads.[1-4, 6-11] An essential aspect for using these probes for single molecule detection is the development of the chemistry that enables site-specific attachment of the probe to the biomolecule of interest. The remaining key challenge is the development of electron spin probes that can be detected at the single molecule level and site-specifically attached to biomolecules. Such a probe could enable electron paramagnetic resonance (EPR) studies of single biomolecules.

EPR is a proven and versatile method of quantifying time scales and the degree of rotational freedom of molecular motion.[12-18] Motion on sub-nanosecond through millisecond timescales can be explored using existing continuous wave and pulsed EPR techniques, [16, 17] offering a potentially powerful enhancement in dynamic range available in studies of individual biomolecules. The power of EPR for measuring structural dynamics at widely varying rates arises from the sensitivity of the EPR spectral lineshape to motion that alters the magnetic resonance frequency. This has the advantage that fluctuations due to motion, characterized by a correlation time $\tau$, will be evident through its impact on linewidth and hence limited by intrinsic spectral linewidth relative to the mechanism responsible for motion dependent broadening, the Zeeman interaction in this case. Fluctuations arising from motion will average the dephasing induced by this interaction leading to a linewidth proportional to $\tau$. The ability to tune this via the Zeeman interaction makes the sub-nanosecond regime accessible. This approach complements imaging approaches such as fluorescence resonance energy transfer (FRET), optical tweezers and magnetic tweezers that measure motion through sequential measurements in the time-domain, and thus are limited for short $\tau$ by their sampling rate. However, the severe sensitivity limitation imposed by conventional inductive detection of the EPR signals of site-specifically attached spin probes, such as a nitroxide, imposes the requirement that measured samples contain $10^{10}$-$10^{15}$ electron spin-labeled molecules [19-21]. In the case of large biomolecular complexes, fabrication of such numbers of molecules can present a serious technical and feasibility challenge. More generally, crucial information on molecular function of biomolecular complexes is masked by the averaging that is inherent to ensemble studies. [1-10] Recent advances using NV diamond ODMR to measure dynamics of adjacent spin labeled proteins [22] and nanomagnets [23] demonstrate this approach for sensitive detection of dynamics. In the complementary approach we present here, the nanodiamond is site specifically attached to a single bio-molecule creating an ESR active, in-situ spin label that can be sensitively detected using NV ODMR for measuring molecular dynamics. This approach to performing EPR on an individual spin-label offers an attractive tool that could enable application of these high-resolution spectral techniques for probing biomolecular structure and dynamics of single molecules.

Here, we take advantage of the nitrogen-vacancy (NV) defect center in diamond that offers an extraordinarily sensitive method of optically detecting the EPR. [24-27] We demonstrate optical detection of EPR from the NV centers in a single-crystal nanodiamond that is attached to a single biomolecule; in particular we report site-specific labeling of double stranded DNA molecules with individual nanodiamonds, the experimental design for optically detecting magnetic resonance from nanodiamond labeled DNA, and experimental measurements of the EPR spectra of these nanodiamond-labeled DNA molecules. The Zeeman shift of the magnetic resonance frequency is sensitive to the orientation of the applied magnetic field relative to the symmetry axis of the NV center making the EPR spectrum sensitive to fluctuations of the orientation of the diamond crystal. This can be used to report on the local flexibility of the biomolecule to which the spin probe is attached, such as is done with EPR studies of nitroxide labeled biomolecules. [15-18] Our single molecule EPR (smEPR) spectra demonstrate that the nanodiamond probe, while attached to a single DNA molecule, explores all available orientations on timescales slower than the spin relaxation time, $T_2$, of the nanodiamond. In this slow motional limit, the spectrum of the single nanodiamond is equivalent to that obtained from a large collection of nanodiamonds because all crystalline orientations will be represented in this static collection; the spectrum in this case is a powder pattern. [12, 13] These studies demonstrate the feasibility of using NV containing nanodiamonds to probe biomolecular dynamics with smEPR and provides a foundation for combining the ability to spectroscopically measure many decades of motional time scales provided by EPR with the power of single molecule methodologies.

## 2. Results

### 2.1. EPR measurements of the NV center in diamond

To perform a continuous wave optically detected magnetic resonance (ODMR) measurement, the NV center is irradiated with a microwave magnetic field, varying its frequency until it matches the energy difference between its $m_s=0$ and $m_s=\pm1$ states thus driving one or both of these transitions. This increases the population of the $\pm1$ states which are able to decay through a non-radiative transition thus reducing the fluorescence from the NV center. [24-27] These magnetic resonance frequencies are shifted in a magnetic field due to the Zeeman interaction with the electronic moment. The magnitude of this shift depends on the orientation of the magnetic field relative to the symmetry axis of the NV center, being largest when the field is parallel to this axis and vanishing when they are perpendicular. In **Figure 1b** we show an ODMR spectrum from an single-crystal diamond which exhibits eight lines, two arising from each of the four orientations of the NV center symmetry axis allowed by the diamond crystalline structure. [24-27] We show the fluorescence quenching as peaks instead of valleys. In this manuscript, we report ODMR measurements of an individual nanodiamond crystal attached to single DNA molecule in which we spectroscopically measure the consequences of rotational motion of the nanodiamond crystal, and hence the molecule, relative to the applied magnetic field.

## 2.2. Experimental Design for ODMR of a Single Biomolecule

An ODMR measurement of a single nanodiamond attached to a DNA molecule requires optical and microwave excitation with optical detection in a controlled aqueous environment. Our experimental setup (**Figure 2**) integrates three key modules: (i) a confocal microscope that optically excites the nanodiamond and efficiently collects the NV fluorescence, (ii) a coplanar waveguide microwave circuit that excites NV electron spins from the $m_s = 0$ to the $m_s = \pm 1$ states, and (iii) fluid circuitry that maintains and controls the DNA molecule-nanodiamond system during the ODMR measurement. **Figure 2a** shows a simplified schematic of the custom-built ODMR microscope used for the smEPR experiment. The optical setup (in grey) and beam path (green pump beam and red NV fluorescence) optically excites and collects the NV luminescence from the labeled sample in the flow cell circuit (in blue). An expanded image of the sample mount (**Figure 2b**) shows the integrated microwave and fluid circuits. Eight fluid-flow channels for housing the single molecule experiments are aligned perpendicular to the Au microwave coplanar waveguide which is photolithographically fabricated directly on the glass surface of the flow cell. The buffer and single DNA molecules labeled with a nanodiamond flow into the channel via a peristaltic pump, which pulls fluid into the channel through drilled holes in the back glass plane of the flow cell. The external magnetic field is applied by a set of rare-earth magnets which deliver a uniform, in-plane magnetic field at the focal spot of the objective. The magnetic field can be rotated 360 degrees in the plane of the sample. Nanodiamond labeled DNA molecules in the co-planar waveguide 'gap' were selected for measurement to ensure sufficient microwave intensity.

### 2.3. Site-specific attachment of a single nanodiamond to a DNA molecule

The application of an NV nanodiamond as a spin probe requires that it be site-specifically attached to a DNA or protein molecule. We took advantage of the biotin-streptavidin-biotin linkage, which has been widely used in previous single molecule measurements. [1, 2, 4, 6, 10] We biotinylated the surfaces of the nanodiamonds (**Figure 3**) prepared by high pressure high temperature (HPHT) methods. HPHT diamonds that have largely sp$^3$ hybridized carbon are much easier to surface functionalize than detonation nanodiamonds, which can contain sp$^2$ (graphitic) carbon surfaces. [28] Following repeated acid reflux cleaning of the nanodiamond, they were incubated in glycidol to introduce alcohol groups on to the surface. The surface was then sequentially treated with *N'*-disuccinimidyl carbonate (DSC), 4,7,10-Trioxa-1,13-tridecanediamine (TTDD), and then NHS-dPEG®-biotin. This resulted in biotin covalently attached to the diamond surface that is sufficiently extended out from the diamond surface for efficient streptavidin binding. The nanodiamonds were then incubated with excess streptavidin, followed by two rounds of centrifugation to remove the unreacted streptavidin.

We used 16μm long Lambda DNA that was labeled at one end with biotin and the opposite end with digoxigenin for the smEPR measurements. The biotin-labeled end of the lambda DNA molecule was attached to a single biotinylated nanodiamond coated with streptavidin, while the digoxigenin labeled end of the Lambda DNA molecules were tethered to the anti-digoxigenin coated glass surface in the smEPR flowcell (**Figure 4a**). [29] Epifluorescence microscopy was used to verify correct linkage between the anti-digoxigenin coated surface and a biotinylated NV nanodiamond (**Figure 4b**). Fluorescence images of SYBRgold (Invitrogen, S11494) labeled single Lambda DNA molecules tethered between the glass surface and a nanodiamond were acquired at two separate emission wavelengths. The images at 570 nm (**Figure 4b**, top) detect the labeled Lambda DNA molecule, while the images at 670 nm (**Figure 4b**, bottom) detect the nanodiamond fluorescence. At a low flow rate of ~0.10 $\mu L/s$, the position of the nanodiamond shifts by 12.4 μm and the fluorophore labeled Lambda DNA tether can be visualized (**Figure 4b**). High flow conditions were avoided to preserve the sample integrity and prevent breaking the tethers off the flow cell surface.

### 2.4. Single Molecule Electron Paramagnetic Resonance

**Figure 5** shows the results of this single-molecule measurement. **Figure 5a** shows the ODMR spectrum of a single static nanodiamond crystal. As discussed above, this results in well-defined peaks where the largest shifts arise from NV centers having the largest projection of applied magnetic field parallel to their symmetry axis. In contrast, a collection of stationary but randomly oriented, nanodiamonds will exhibit a distribution of shifts because all possible orientations are represented (**Figure 5b**); this result is a powder spectrum. The uniform intensity of the spectrum reflects the uniform probability of finding an NV center having any particular orientation relative to the applied field. The spectrum cuts off at a maximum frequency associated with the subset of NV centers whose axes are parallel to the applied field, thus the magnitude of the cutoff is equal to the full Zeeman shift in the particular applied magnetic field.

A similar spectrum results from a single nanodiamond crystal if it dynamically rotates through all possible orientations during the acquisition of the spectrum. **Figure 5c** shows the ODMR spectrum obtained from the single nanodiamond tethered to the end the DNA molecule. The close similarity to the powder spectrum confirms our expectation that the nanodiamond probe rotates isotropically, and hence freely, through all possible orientations. It is also consistent with the fact that, due to its ~100 nm diameter, its characteristic rotational fluctuation timescale, $\tau_R$, is slow compared to the characteristic transverse spin relaxation time $T_2$, typically in the range 0.25-1.4 μs or longer for NV centers in nanodiamond samples. [30] For smaller nanodiamonds having $\tau_R$ smaller than $T_2$, the short correlation time would lead to motional averaging, that is, a reduction of the net effect of motion on the evolution of the NV spin under the influence of the applied magnetic field leading to a reduction of the width of the resonance line and hence a spectrum with sharper lines.

The ODMR spectra of both the powder and the single nanodiamond tethered to the DNA molecule were measured at three different applied fields: 0 Gauss, 19 Gauss and 32 Gauss. The solid lines show a comparison of our simulation of the spatially-averaged powder dispersion spectrum for nanodiamonds that was fit to both data sets. The details of this data fitting can be found in the materials and methods section. The field values extracted from these fits are 0 G (red) (fixed), 18.7 ± 0.05 G (blue), and 32.6 ± 0.09 G (green) for the powder spectra, 0 G (red) (fixed), 19.0 ± 0.26 G (blue), and 32.1 ± 0.28 G (green), for the smEPR spectra. The fitted values agree well, within the error, with our determinations of the fields applied by means of a set of moveable rare-earth magnets.

## 3. Discussion and Conclusions

The combination of optical detection of EPR within a buffered biocompatible solution with the site-specific attachment of an NV nanodiamond to a DNA molecule demonstrates the feasibility of EPR measurements on a single spin-labeled biomolecule. Given the wide use of ensemble EPR measurements of spin-labeled biomolecules,[15-18] the realization of single molecule EPR helps enable the development of a broadly applicable methodology for investigating single biomolecular structures and their dynamics. Furthermore, our attachment strategy of biotin-streptavidin-biotin for site-specific labeling of a nanodiamond to a DNA molecule is compatible with RNA , proteins, and lipid vesicles.[1-4, 6, 8-10] This suggests that our approach to NV diamond site-specific labeling can be used with a wide variety of biomolecular systems.

Given the potential of this single molecule methodology, we consider the ultimate scope of this continuous wave (CW) EPR approach. There are two key time scales that influence this application: (i) the mechanical rotation correlation time of the spin label, $\tau_R$, and (ii) the ensemble spin lifetime, $T_2$.[30] For the case where $\tau_R \gg T_2$, the CW EPR spectrum depends on the range of rotational freedom of the nanodiamond label. As the nanodiamond label converts from unrestricted to highly constrained rotational motion, the EPR spectrum will convert from a powder pattern to a spectrum with eight peaks, as we observed in this study (Fig. 5). These observations suggest that analysis of the nanodiamond EPR spectrum could be used to determine the rotational freedom of the biomolecular region around the site of nanodiamond attachment, as is used in ensemble CW EPR measurements.[15-18]

For the case where $\tau_R \ll T_2$, the mechanical rotation of the nanodiamond will motionally narrow the spectrum as is observed in other spin-labelled systems.[15-18] The nanodiamond $\tau_R$ can be estimated by considering the rotational correlation time of a sphere, $4\pi r^3 \eta / 3k_B T$, where r is the radius, $\eta$ is the viscosity of water ($10^{-3}$ Pa sec) and $k_B T$ is the thermal energy unit at room temperature. The data in **Figure 5** represents a measurement on a ~100 nm nanodiamond, which implies a $\tau_R \approx 100$ μsec. Given that the nanodiamond $T_2$ is 0.25 – 1.4 μs,[30] the measurements reported here are in the limit $\tau_R \gg T_2$, so consistent with our measured powder pattern EPR spectrum. However, measurements where $\tau_R \leq T_2$ appear achievable. Reduction of the nanodiamond size to reported values of 10 nm will reduce $\tau_R$ to about 0.1 μsec, which is less than the reported nanodiamond $T_2$. [30] In addition to reducing $\tau_R$, $T_2$ can be increased by improving the purity of the nanodiamond. Spin lifetimes as long as 0.7 msec have been reported for highly pure nanodiamonds in aqueous environments.[31] These results combine to suggest that partial motional narrowing could lead to modifications of the single nanodiamond EPR spectrum, providing information about the local biomolecular dynamics of the region surrounding the site of nanodiamond attachment.

Beyond these CW applications to individual biomolecules, many of the advances described above have important implications for related techniques and experimental geometries. For example, the attachment chemistry developed for the nanodiamonds can be used in conjunction with recent experiments demonstrating *in vivo* measurements[32] to achieve targeted labeling of cellular structures. That same study demonstrated *in vivo* measurements on NV nanodiamonds. [32] A separate study demonstrated CW EPR of a single nanodiamond within an optical trap.[33] The ability to site-specifically attach a biomolecule to a nanodiamond demonstrated here provides a key step toward the integration of force spectroscopy via optical trapping[2, 3, 6, 8] and EPR spectroscopy.

Finally, we note that the CW EPR techniques employed in this study provide a foundation for the development of more sophisticated pulsed EPR techniques in much the same way that ensemble CW EPR studies paved the way for pulsed studies of large ensembles. A promising next step is to apply pulsed ODMR techniques to this measurement which will enable substantially improved spectral resolution.[34] In combination with smaller nanodiamond labels, this could enable quantitative measurement of both motional timescales and the rotational freedom of the single molecules to which the probe is attached.

## 4. Experimental Section

*Nanodiamond Sample Preparation*: The NV nanodiamonds (Van Moppes SYP0.09) are synthesized from a high pressure, high temperature diamond with a nitrogen impurity content of $n_N$ ~200 ppm. The diamond is milled down via micro-fracturing into mono-crystalline nanodiamonds with a size distribution of 10-200 nm in diameter. The nanodiamond is then irradiated (Prism Gem) with a 1.5M eV, $3.48 \times 10^{18}$/cm$^2$ per hour electron beam for three hours to create a vacancy density of $n_V$ ~59 ppm. The nanodiamond is then annealed at 900 °C for 3 hours in 96% Ar and 4% H$_2$ to bring the nitrogen impurities and vacancies adjacent to each other to create NV centers which typically leads to an NV center density of approximately 30 ppm.[35-38]

Next, the nanodiamonds are cleaned to remove any graphitic residue from the surface. The removal of the graphite both prepares the surface of the diamond for biochemical functionalization[28, 39] and also reduces damping of the fluorescence of the NV centers near to the surface of the nanocrystal due to surface effects.[30, 39, 40] The nanodiamond powder is acid-cleaned under reflux. First it is boiled in a 9:1 mixture of H$_2$SO$_4$ (98%) and HNO$_3$ (70%) at 90 °C for 3 days. Next, it is washed in deionized water and then resuspended in a new 9:1 acid mixture and boiled for 1 day. Then, it is washed and resuspended in 0.1 M NaOH and boiled at 90 °C under reflux for 2 hours. Finally, it is washed and resuspended in 0.1 M HCl and boiled at 90 °C for another 2 hours. The first two acid-washing steps in H$_2$SO$_4$ and HNO$_3$ are then repeated. And lastly, it is triple rinsed with deionized water and ready for surface functionalization.[28, 30, 39-44]

*Nanodiamond Surface Biotinylation*: The acid-reflux cleaning of the nanodiamond prepares the surface with terminal carboxyl groups.[28, 39, 42-44] The nanodiamond is then reacted with glycidol (Sigma-Aldrich, St. Louis MO) to create a hydroxyl terminated surface.[45, 46] The diamonds were then rinsed three times with dimethylacetamide (DMAC) and resuspended in 100uL of the same solvent. This was diluted into 900uL of dimethylformamide (DMF) containing 100 mM *N'-disuccinimidyl carbonate* (DSC, EMD

Millipore, Billerica, MA) and allowed to react for 2 hours at room temperature. Unreacted DSC was removed by washing three times in DMAC then quickly rinsing once in cold phosphate-buffered saline containing 0.05% Tween 20 (PBST). The NHS-activated diamonds were resuspended in PBST having a 1% solution of 4,7,10-Trioxa-1,13-tridecanediamine (TTDD, Sigma-Aldrich St. Louis MO). Reaction was incubated for 2 hours at room temperature and unreacted TTDD was removed by exhaustive rinsing with PBST. The resulting amine-PEG$_3$-functionalized diamonds were finally reacted with 0.22 μM/mL NHS-dPEG®-biotin (Quanta Biodesign, Powell, OH) for 2 hours, triple rinsed in PBST and finally suspended in PBST for the experiment. This process is illustrated in Figure 3a.

*Lambda DNA preparation:* Lambda DNA functionalized with biotin and digoxigenin was prepared using commercially available linear lambda DNA (New England Biolabs) which contains a 12 nucleotide 5' overhang on either end. Biotin was attached, by ligating a synthetic oligonucleotide (Operon) containing a 3' biotin (5'-AGGTCGCCGCCC-biotin-3'), while the digoxigenin was attached by ligating a synthetic oligonucleotide (Operon) containing a 3' digoxigenin (5'-GGGCGGCGACCT-Dig-3') to the lambda DNA. The ligations were done simultaneously with a 100 fold molar excess of each oligonucleotide in 1X T4 Ligase buffer, 1X BSA and 400 units of T4 Ligase (New England Biolabs). This reaction was allowed to proceed for 4 hours at room temperature, then incubated at 65°C for 20 minutes to heat kill the ligase enzyme. Functionalized lambda DNA was purified from excess oligonucleotide ends via a Microspin G-50 column (GE Healthcare Lifesciences).

*DNA-Nanodiamond Attachment Chemistry:* In preparation for the DNA attachment to the flow cell surface, the flow-cell cover slip is Piranha cleaned (2:1, $H_2SO_4$ to $H_2O_2$) and then treated with 3-Aminopropyl-Triethoxysilane (MP Biomedicals 154766) to terminate the surface with Amine groups. Next, the surface is coated with Glutaraldehyde ($OCHC_3H_6HCO$, Electron Microscopy Sciences 16320) and then, lastly, terminated with Antidigoxygenin antibody (Roche Diagnostics 11333089001).[29] The DNA molecule is bound to the glass surface of our flow cell using an Antidigoxygenin-Digoxygenin antibody attachment. A schematic of this chemistry is given in Figure 3b.

To create the single DNA-nanodiamond attachments on the flow cell surface, first the biotinylated NV nanodiamond is pipetted into 1 mg/mL Streptavidin (Sigma S4762-1MG). After a 10 minute incubation period, the excess Streptavidin is removed with two washes in phosphate buffered saline (PBS). The Streptavidin-coated NV nanodiamond is resuspended in PBST and 0.2 mg/mL bovine serum albumin (BSA). The Lambda DNA is gently pipetted into the sample and allowed to incubate for 15 minutes. The sample is finally pumped into the prepared flow cell and allowed to incubate for 20 minutes, allowing ample time for the Digoxygenin tag on the DNA molecule to attach to the flow cell surface. The cell is washed with PBST-BSA to remove any excess DNA.

*Custom-designed Confocal Microscope:* As shown in **Figure 2a**, the sample is pumped with a continuous wave laser (532 nm, 300 mW DPSS laser, SNOC electronics) through the objective (Nikon Plan Fluor 100x oil-immersion, 1.3 NA), and the subsequent photoluminescence from the NV centers is collected back through the objective and directed onto an avalanche photodiode (APD, Pacific Silicon Sensor AD500-8-S1BL). The APD signal is processed through a lock-in amplifier (Model 7265 DSP Lock In from Signal Recovery) at 400 Hz, which matches the amplitude modulation frequency of the microwave source (Agilent 8648C). The continuous microwave radiation is delivered to the sample through coaxial cables connected to the sample mount. The microwave and laser radiation powers are kept low to prevent sample heating and preserve sample integrity. Our DNA-nanodiamond samples demonstrated stability over many hours of continuous excitation during our measurements. The external magnetic field is applied via a set of rare earth magnets that can be rotated $360^\circ$ in the sample plane.

*Powder spectrum modeling and fitting:* To analyze the single molecule EPR spectrum, we simulated the spectrum and fit the simulation to the data, shown as the solid lines in

**Figure 4a-b**. From these fits we can extract the applied field, strain, and resonance linewidth of the NV centers. The NV center is well described by the Hamiltonian[26]

$$H = D\left(S_z^2 - \frac{2}{3}\right) + g_e \mu_B \boldsymbol{B} \cdot \boldsymbol{S} + E\left(S_x^2 - S_y^2\right) \tag{1}$$

where $D$ = 2.87 GHz is the zero-field splitting, $g_e$ is the electron g-factor, $\mu_B$ is the Bohr magneton, $B$ is an externally applied magnetic field, $E$ is the strain, and $S_x$, $S_y$, $S_z$ are the spin-1 Pauli matrices. Since carbon-13 is 1.1% abundant, some NV centers will also experience a strong hyperfine interaction with the carbon-13 nuclear spin, which adds an additional term to the Hamiltonian, $\boldsymbol{S} \cdot \boldsymbol{A} \cdot \boldsymbol{I}$, where $A$ is the hyperfine tensor and $I$ is the carbon-13 nuclear spin. Diagonalization of the Hamiltonian yields the allowed eigenstates, which are sensitive to the direction of the applied magnetic field. Each NV center is defined along one of diamond's four crystallographic axes, and the zero-field splitting term quantizes the NV center's z-component of spin along this axis. The eigenstates depend upon the angle between the NV axis and the applied magnetic field, which is important for fitting the powder of nanodiamonds, in which there are NV centers randomly oriented in any direction.

We generate the powder pattern spectra shown in **Figure 4** by using the conventional formula described by Abragam.[13] We find the energy splitting of the two allowed spin transitions, i.e. between the 0 and +/- 1 spin states, at a particular angle. We use a Lorentzian lineshape to describe the resonance, which yields a simulated resonance spectrum for a single nanodiamond. We then find the spectrum averaged over all angles in order to simulate the full powder spectrum. We apply this method to simulate both the static spectrum of many nanodiamonds dispersed on a surface, and the dynamic spectrum of a single nanodiamond attached to a DNA strand. Our experiment records the spectrum of the attached nanodiamond as it varies it's orientation in time, exploring all orientations, such that a powder spectrum is a good approximation.

We then fit the simulated spectrum to the data to extract the relevant physical parameters. We first fit the zero-applied field powder spectrum data to extract the linewidth and strain, which causes the prominent dip in the center of the spectrum. The spectrum is the superposition of two spectra, NV centers without $^{13}$C interactions, and NV centers with $^{13}$C interactions which cause the bumps on the edges of the spectra. The relative amplitudes for these two populations is also a fit parameter. We then fix linewidth, strain, and relative amplitude of the $^{13}$C-interacting population, and only allow the applied magnetic field to be a fitting parameter for the spectra with applied fields. For the powder spectra in **Figure 4**, the fits give estimated parameter errors which are less than a Gauss for the three spectra with applied magnetic field.

**Acknowledgements**

We acknowledge primary support from the Center for Emergent Materials: an NSF MRSEC under award number DMR-1420451, and partial support from the ARO through award W911NF-12-1-0587 and the NIH through award GM083055.

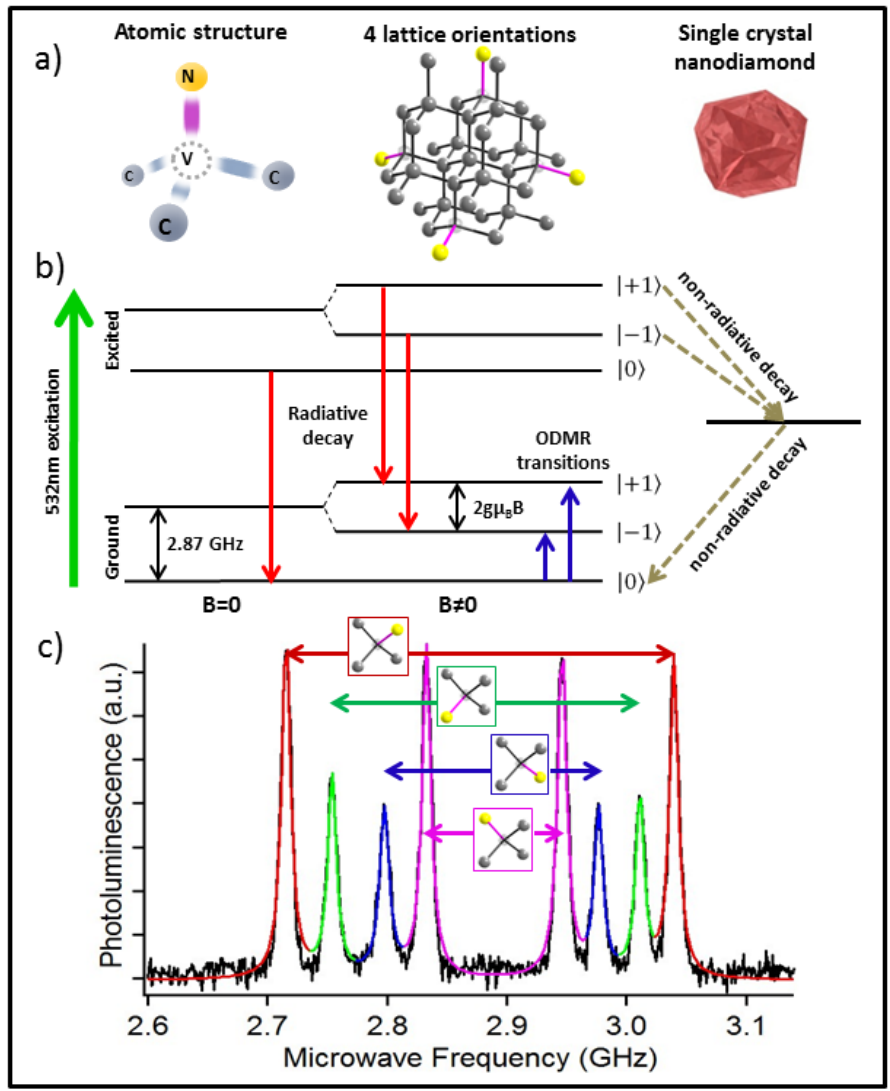

**Figure 1. a)** A schematic of the NV center atomic structure (left), composed of a nitrogen impurity adjacent to a vacant site in the diamond tetrahedral lattice. The NV center axis can adopt any of 4 allowed orientations in the diamond lattice (middle). Our NV nanodiamonds are single-crystal and range from 10-200 nm in diameter (right). **b)** A simplified schematic of the electronic structure of the NV center. **c)** An optically detected magnetic resonance spectrum of a collection of NV centers in a diamond crystal exhibiting the Zeeman splitting of the magnetic resonance peaks by an applied magnetic field (shown by dark blue arrows) and the dependence of the splitting on the orientation of the field relative to the NV axis.

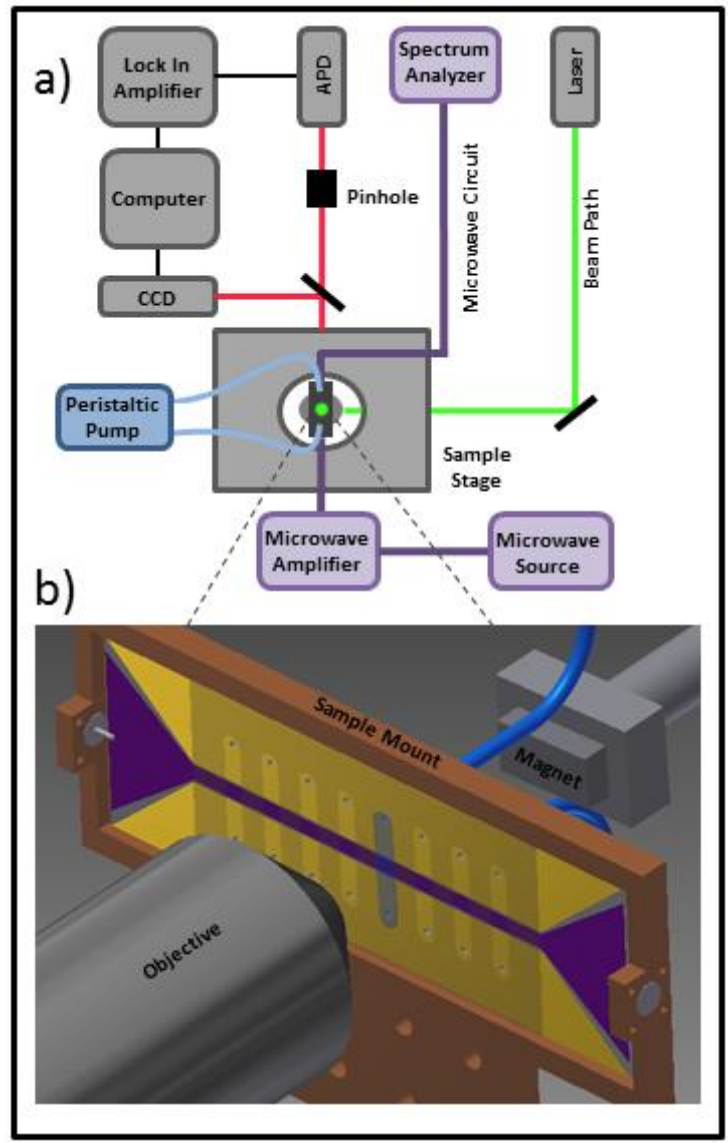

**Figure 2. a)** A simplified schematic of the custom-built confocal microscope. The optical components (in grey) are tightly integrated with the microwave circuit (components in purple) and flow cell circuit (in blue). **b)** A zoomed in view of the integrated microwave and fluid circuits at the sample, illustrating eight flow channels for housing the in vitro single molecule experiment, running perpendicular across the microwave coplanar waveguide. The microwave circuit path is highlighted in red and the fluid circuit is in blue.

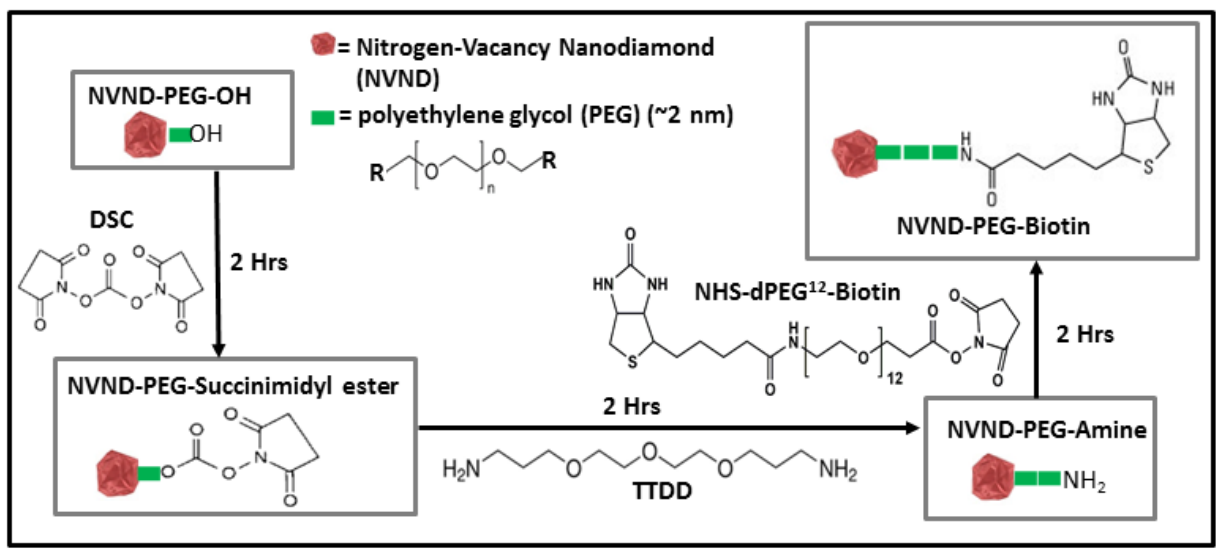

**Figure 3.** The work flow of the NVND biotinylation chemistry protocol.

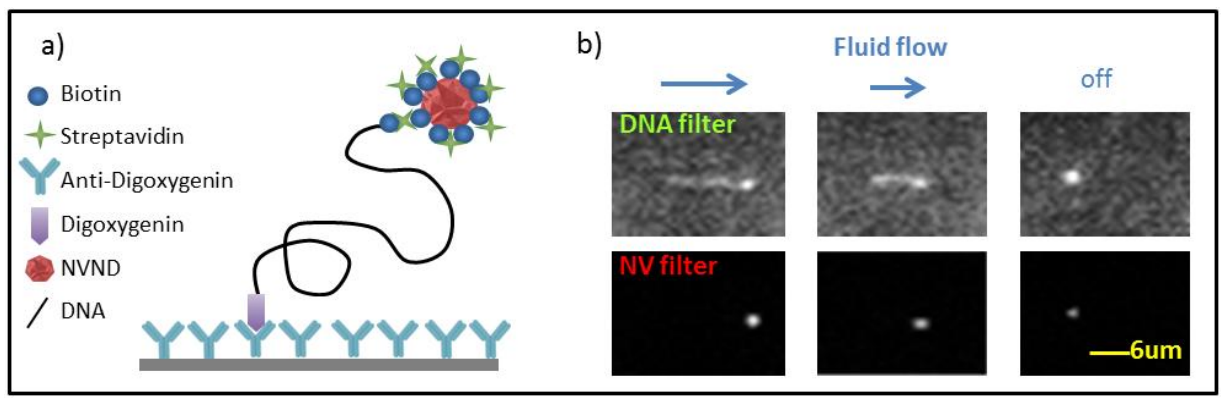

**Figure 4.** The attachment of a single nanodiamond to a single DNA molecule. **a)** A schematic representation showing the binding between streptavidin, anti-digoxygenin antibodies, individual, dual-labeled (digoxygenin, biotin) Lambda DNA molecule and biotinylated fluorescent nanodiamonds. The assembly occurs on the surface of the glass coverslip. **b)** An epifluorescence microscope image of the attached nanodiamond. On the top is an image of the surface through a 570 nm band pass filter. On the bottom is the same surface as seen through a 670 nm bandpass filter. The NV diamond fluorescence will pass through the 670 nm filter and the SYBR gold-dyed DNA molecules will emit through the 570 nm bandpass filter only.

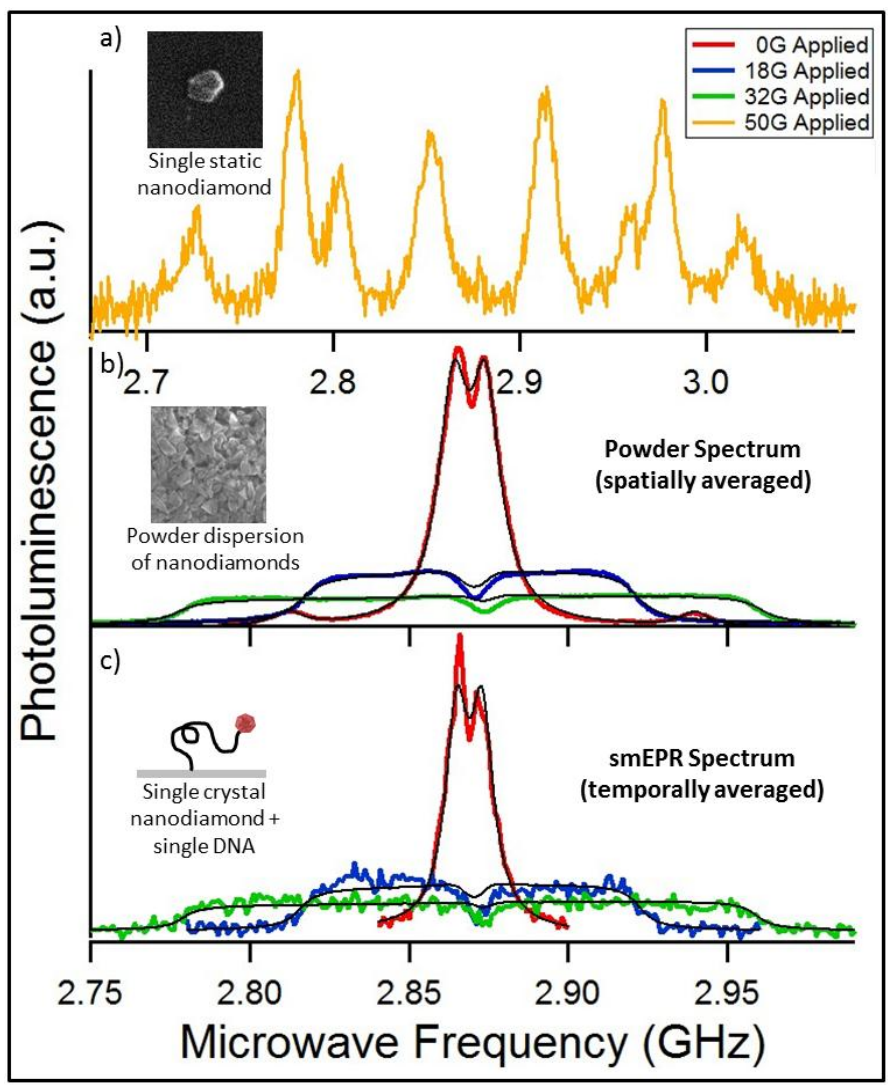

**Figure 5.** Optically detected magnetic resonance (ODMR) spectra. **a)** ODMR spectrum of a single static nanodiamond. The eight distinct peaks arise from NV centers oriented in four directions allowed by the nanodiamond crystal lattice. **b)** NV diamond powder dispersion, collected from a large population of randomly oriented nanodiamond crystals (top inset) under three different applied external fields: 0 G (red), 18.7 G (blue), and 32.6 G (green). The uniform intensity of this spectrum reflects the uniform probability of finding an NV center having any particular orientation relative to the applied field. The field values are extracted from the fit for each spectrum (black), based on our model and are in agreement with our field calibration. **c)** The smEPR spectrum of a nanodiamond crystal attached to a single DNA molecule (bottom inset) under three similar applied external fields: 0 G (red), 19.0 G (blue),

and 32.1 G (green), again, extracted from the fit (black) for each spectrum, also based on our model. The close similarity to the powder spectrum confirms our expectation that the nanodiamond probe rotates isotropically, and hence freely, through all possible orientations on a timescale that is slow compared to the characteristic transverse spin relaxation time $T_2$ for NV centers.